\documentclass[preprints,tutorial,accept,pdftex,moreauthors]{mdpi} 
\firstpage{1} 
\pubvolume{1}
\issuenum{1}
\articlenumber{0}
\pubyear{2024}
\copyrightyear{2024}
\datereceived{ } 
\daterevised{ } 
\dateaccepted{ } 
\datepublished{ } 
\hreflink{https://doi.org/} 
\pdfoutput=1 

\usepackage[arrowdel]{physics}
\usepackage{subcaption}
\Title{A tutorial on the use of physics-informed neural networks to compute the spectrum of quantum systems}

\TitleCitation{A tutorial on the use of physics-informed neural networks to compute the spectrum of quantum systems}

\Author{Lorenzo Brevi $^{1}$\orcidA{}, Antonio Mandarino $^{1}$\orcidB{} and Enrico Prati$^{1}$\orcidC{} }

\AuthorNames{Lorenzo Brevi, Antonio Mandarino and Enrico Prati}

\AuthorCitation{Brevi, L.; Mandarino, A.; Prati, E.}

\address[1]{%
$^{1}$ \quad Department of Physics Aldo Pontremoli,Università degli Studi di Milano, Via Celoria 16, 20133 Milano, Italy}

\abstract{Quantum many-body systems are of great interest for many research areas, including physics, biology and chemistry. However, their simulation is extremely challenging, due to the exponential growth of the Hilbert space with the system size, making it exceedingly difficult to parameterize the wave functions of large systems by using exact methods. Neural networks and machine learning in general are a way to face this challenge. For instance, methods like Tensor networks and Neural Quantum States are being investigated as promising tools to obtain the wave function of a quantum mechanical system.
In this tutorial, we focus on a particularly promising class of deep learning algorithms. We explain how to construct a Physics-Informed Neural Network (PINN) able to solve the Schr\"odinger equation for a given potential, by finding its eigenvalues and eigenfunctions. This technique is unsupervised, and utilizes a novel computational method in a manner that is barely explored. PINNs are a deep learning method that exploits Automatic Differentiation to solve Integro-Differential Equations in a mesh-free way. We show how to find both the ground and the excited states. The method discovers the states progressively by starting from the ground state. We explain how to introduce inductive biases in the loss to exploit further knowledge of the physical system. Such additional constraints allow for a faster and more accurate convergence. This technique can then be enhanced by a smart choice of collocation points in order to take advantage of the mesh-free nature of the PINN. The methods are made explicit by applying them to the infinite potential well and the particle in a ring, a challenging problem to be learned by an Artificial Intelligence agent due to the presence of complex-valued eigenfunctions and degenerate states}


\newcommand{\cor}[1]{{ #1}}

\begin{document}

\section{Introduction}
The deep learning (DL) approach is one of the most successful machine learning (ML) paradigms, carrying outstanding results on several complex cognitive tasks \cite{DeepLearningNature, DeepLearningReview2}. Training a Deep Learning model, however, usually requires an enormous amount of data. Even when the dataset is large enough to allow for effective training, such data will be finite and noisy, inevitably leading to issues such as overfitting. However, when addressing the dynamics of a physical system there is often prior knowledge about its behavior and the laws that shape it, usually in the form of a system of Integro-Differential Equations. 

Physics-informed Neural Networks \cite{RAISSI2019686} are a promising tool to discover and address the parametrization of a system governed by Partial Differential Equations (PDEs) or Integro-Differential Equations \cite{YUAN2022111260}. This rather new family of models has already shown success in a plethora of different fields \cite{review2021, review2024}, thanks to their ability to learn the implicit solution of a PDE even when given little to no data \cite{brevi2024addressing}. It allows to build models able to more efficiently simulate those systems and even to discover new physics \cite{Raissi20201026}. Furthermore, integrating the physical knowledge of the system in the training process allows to prevent overfitting. 
They have been widely applied in fluid dynamics \cite{PhysRevFluids.4.034602, Raissi20201026, PhysRevFluids.6.073301} where, among many other things, they have been used in predictive large-eddy-simulation wall modeling, to obtain a model with improved extrapolation capabilities to flow conditions that are not within the training data. Unknown physics has been learned by improving on empirical models for viscosity in non-Newtonian fluids. There, the PINN model removes an unphysical singularity present in the empirical one. In astrophysics \cite{PhysRevD.107.064025}, they have been used to compute the first quasinormal modes of the Kerr geometry via the Teukolsky equation. The resulting model allows to compute the oscillation frequencies and the damping times for arbitrary black hole spins and masses. They have also been employed in scattering in composite optical materials \cite{Chen202011618}, reconstruction of top quark kinematic properties \cite{PhysRevD.107.114029}, and to compute the electronic density in complex molecular systems \cite{PhysRevLett.125.206401}. However, in all the aforementioned cases the physics has just been used as a way to augment experimental data. The approach explained in this paper is rather different. Here the aim is to solve the Schr\"odinger's equation for a system given its governing potential, in an unsupervised way.

Previously, some of us have developed both supervised \cite{maronese2022quantum,moro2023anomaly,corli2023max} and unsupervised \cite{rocutto2021quantum} machine learning methods, including supervised quantum machine learning to address ground state classification \cite{lazzarin2022multi, LMG_grossi, Monaco_ANNNI} of a physical system. The quantum anharmonic oscillator, typically used to describe the quantum dynamics of the Josephson junctions in superconducting qubits has been addressed despite being non-integrable \cite{brevi2024addressing}. Following this latter approach we aim to formalize the unsupervised solution of Schr\"odinger's equation \cite{PINNschrod} and explore the potential of Physics-Informed Neural Networks (PINNs) for solving Partial Differential Equations (PDEs) as a paramount representative application to nontrivial physical phenomena. 


This tutorial explains how to apply unsupervised PINNs to Schr\"odinger's equation  of a quantum mechanical system in a self-consistent way, in order to find both the ground state and the excited states of it. As illustrative embodiment of this method the infinite potential well and the particle in a ring are solved \cite{AIforQT}. This tutorial is divided as follows: in Section \ref{sec: tutorial} we explain how to build the PINN and in particular its loss function. Then in Section \ref{sec: evaluation} we show how to evaluate the training of the model to tune its hyperparameters in a more informed way. Lastly, in Section \ref{sec: examples} we solve the two systems that exemplify the method.

\section{Materials and Methods} \label{sec: tutorial}
\begin{figure}[H] 
    \centering
    \includegraphics[width=1.0 \linewidth]{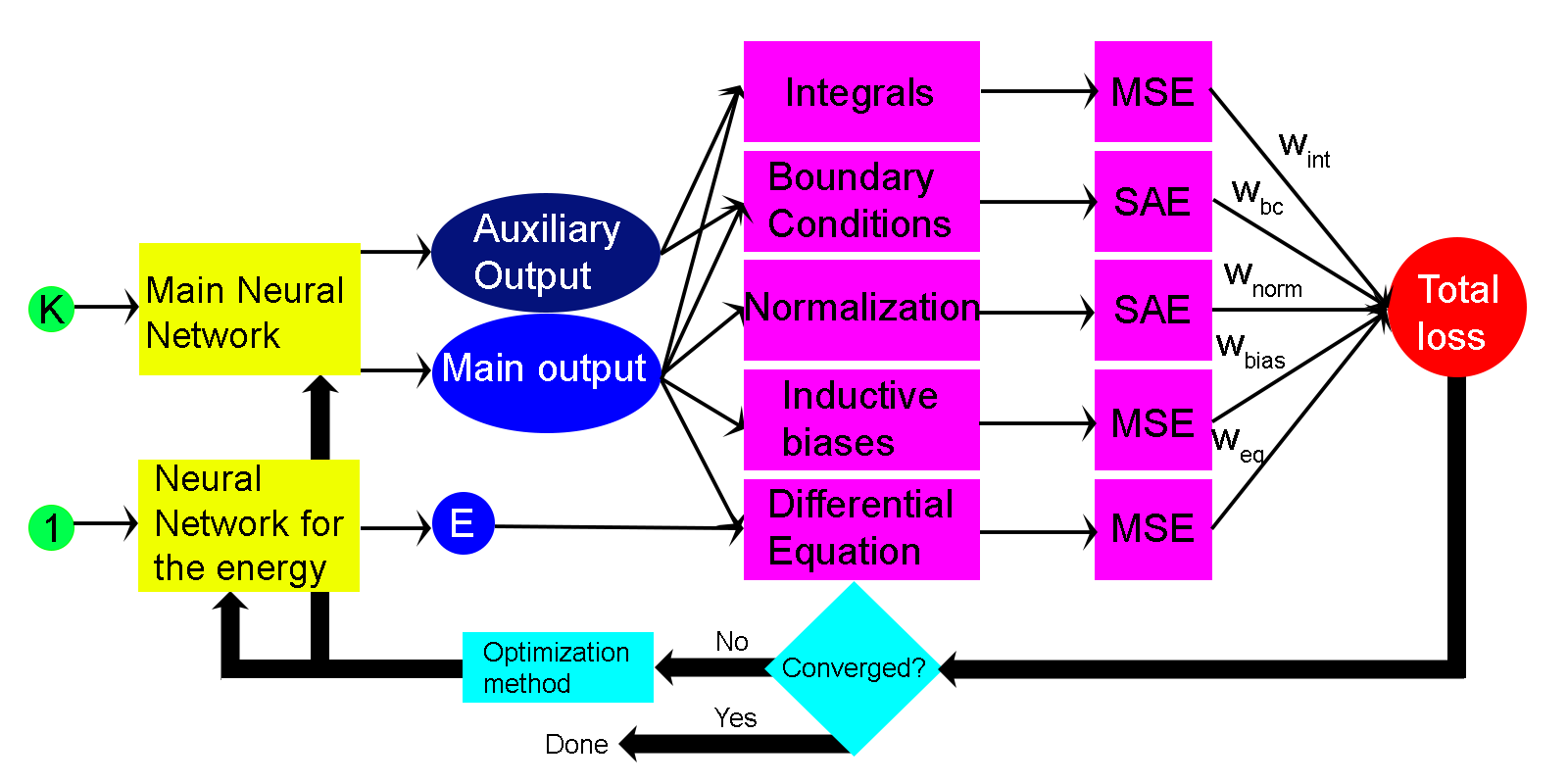}
    \caption{\cor{Graphical representation of the approach followed in this paper. The circles represent variables, the rectangles represent operations. The variables are also represented by primary colors, while the methods are represented by the secondary colors obtained by combining the colors of their input and their output. In green there are the neural network inputs. In yellow there are the two neural networks, the main one that computes the eigenstates and the one that computes the energy. In blue there are the outputs of the neural networks: the auxiliary outputs, the main outputs which is the eigenfunction and the eigenvalue $E$. In purple there are the operations needed to calculate the losses. The losses are the integral loss, the boundary condition loss, the normalization loss, the inductive biases and the differential equation loss. SAE is the Sum of Absolute Errors, SSE is the Sum of Squared Errors. In red there is the final loss. It is calculated by summing the partial losses weighted by empirically adjusted hyperparameters $w_{int}$, $w_{bc}$, $w_{norm}$, $w_{bias}$ and $w_{eq}$. In cyan there are the criteria and the optimization method.}}
    \label{fig:graph}
\end{figure}

In this Section, we explain how to build a PINN and how to decide design the loss function and to apply heuristics to optimize the network's hyperparameters. First, Section \ref{sec: NN_generic}  contains in a concise form the required background information on artificial neural networks and deep learning. Then in Section \ref{sec: architecture} we give a detailed explanation of what a PINN is and how to build it, and the auxiliary output method used to calculate the normalization is explained. Afterwards, in Section \ref{sec: preliminaries}, we explain how to analyze the problem to understand the best possible way to encode the physics in the loss function of the PINN. After that in Section \ref{sec: constraint} we show how to enforce the physical constraints of the system, such as normalization and boundary conditions. After those Sections the reader should have gained a picture of the main losses needed to find an eigenstate for the given potential. More work is, however, required to find the corresponding eigenvalue. In Section \ref{sec: energy} we show how to find the energy by solving the equation in a self-consistent way. Next, the problem of inducing the PINN to converge to a specific eigenstate is addressed. In particular, Section \ref{sec: orthogonality} demonstrates how to build a series of PINNs where the first predicts the ground state of the system while each subsequent network predicts the next state, in order of increasing energy. At this point, the problem has been set up. The next Sections show how to build the best possible computational framework to solve it: in Section \ref{sec: weights} it is explained how to set the best weights, which are given special attention since those are the main hyperparameters to tune for the PINN. Then in Section \ref{sec: training points} we introduce a method to choose the best set of training points, and in Section \ref{sec: domain} how to choose the computational domain. Lastly, Section \ref{sec: NN} gives some pointers on how to best build the actual network, with the number of neurons, hidden layers, activation function, and loss metrics. This pipeline is summarized in Figure \ref{fig:graph}

\subsection{Neural Networks and Deep Learning} \label{sec: NN_generic}

This Section gives a broad qualitative overview of the basic tools needed for this tutorial, namely deep neural networks. The information here has been largely taken from Refs. \cite{Kruse2022} and \cite{Alzubaidi2021ReviewOD}.
    (Artificial) neural networks are information processing systems, whose structure and operation principles are inspired by the nervous system and the brain of animals and humans. They consist of a large number of fairly simple units, the so-called neurons, which are working in parallel. These neurons communicate by sending information in the form of activation signals, along directed connections, to each other. 
 Deep Learning algorithms face a series of challenges to train them effectively. For instance, one of the most common issues is the problem of overfitting. With a larger number of hidden neurons, a neural network may adapt not only to the regular dependence between outputs and inputs we try to capture, but also to the accidental specifics (and thus also the errors and deviations) of the training data set. This is usually called overfitting. Overfitting will usually lead to the effect that the error a neural network with many hidden neurons yields on the validation data\footnote{A second dataset whose labels are known that is not employed in training but only to get a more accurate estimate of the error of the network} will be greater than the error on the training data. This is due to the fact that the validation data set is likely distorted in a different fashion, since the errors and deviations are random. Furthermore, while DL methods are known to be extremely effective when immense amounts of data are available, this availability is not a given in all circumstances. For instance, the data might be costly to obtain. Or it might be tied to the observation of some phenomenon that cannot be replicated at will and is not common enough to allow the training of a DL model that relies on big data. PINNs aim to \cor{address} both the problem of data availability and mitigate overfitting by making use of the underlying physics of the system.

Mathematically, an artificial neural network is a (directed) graph $G = (U,C)$ whose vertices $u \in U$ are called neurons or units and whose edges $c \in C$ are called
connections. The set $U$ of vertices is divided into the set $U_{in}$ of input neurons, the set $U_{out}$ of output neurons and the set $U_{hid}$ of hidden neurons. These sets must fulfill the following conditions:
\begin{equation}
    \begin{aligned}
        U = U_{in} \cup U_{out} \cap U_{hid}\\
        U_{in} \neq \emptyset, \; U_{out} \neq \emptyset\\
        U_{hid} \cap \left(U_{in} \cup U_{out} \right) = \emptyset
    \end{aligned}
\end{equation}

Each connection $(u, v) \in C$ carries a weight $w_{uv}$ and to each neuron $u \in U$ three real-valued quantities are assigned: the network input $net_u$, the activation $act_u$
and the output $out_u$. In addition, each input neuron $u \in U_{in}$ has a fourth quantity, the external input $ext_u$. Each neuron $u \in U$ also possesses three functions:
\begin{equation}
    \begin{aligned}
        \textrm{the network input function}& \;\;\;\; f_{net}^{(u)}: 	\mathbb{R}^{2|pred(u)|+k_1(u)} \to 	\mathbb{R} \\
        \textrm{the activation function}& \;\;\;\; f_{act}^{(u)}: 	\mathbb{R}^{k_2(u)} \to 	\mathbb{R} \\   
        \textrm{the output function}& \;\;\;\; f_{out}^{(u)}: 	\mathbb{R} \to 	\mathbb{R} \\ 
    \end{aligned}
\end{equation}
with which the network input $net_u$, the activation $act_u$ and the output $out_u$ of the
neuron u are computed. $k_1(u)$and $k_2(u)$ depend on the type and the number of arguments of the functions. The neurons are divided into input, output, and hidden neurons in order to specify
which neurons receive input from the environment (input neurons) and which emit
output to the environment (output neurons). The remaining neurons have no contact
with the environment (but only with other neurons) and thus are “hidden.” To operate, for each neuron of the neural network The network input function $f^{(u)}_{net}$ computes the network input $net_u$ from the inputs
$in_{uv1}$, $\dots$ , $in_{uvn}$ (which correspond to the outputs $out_{v1}$, $\dots$ , $out_{vn}$ of the predecessors
of the neuron $u$) and the connection weights $w_{uv1}$, $\dots$ ,$w_{uvn}$. This computation may be influenced by additional parameters. The activation
function $f^{(u)}_{act}$ computes the new activation $act_u$ of the neuron $u$. Finally, the output
of the neuron $u$ is computed by the output function $f^{(u)}_{out}$ from its activation. Note that the choice of the activation function in particular can greatly influence the convergence of the network. The network's parameters can be determined by training it utilizing Gradient Descent and Backpropagation. The idea is that we aim to train the network's parameters so that they minimize a function, the loss function, that quantifies the distance from the network's output and the desired output. By choosing a sufficiently smooth activation function (i.e. the logistic function or $tanh$) it is possible to compute the gradient of the loss with respect to the parameters. Then, by changing the parameters in the inverse direction as this gradient we can minimize the loss function. This also means that we need a differentiable function that has a non-vanishing derivative in at least a sufficiently large region, where it provides direction information. However, computing this gradient for a deep neural network is not trivial. It requires a method to compute the derivatives of the outputs (and therefore of the loss) with respect to the weights of both the ouptut layers and the hidden layers. This method is called error backpropagation. It exploits the chain rule to obtain a layer-wise recursion formula for computing the derivatives of the
neurons of the hidden layers. Note that this method can easily be extended to compute the derivative of the outputs with respect to the inputs, which is crucial for training PINNs.

\subsection{Materials} 
Due to the pedagogical nature of this article, we decided to not use examples that require expensive computational resources. 
Therefore, all the results in the paper are reproducible using a laptop with an NVIDIA GeForce RTX 3060 Laptop GPU.

\subsection{Physics-Informed neural networks} \label{sec: architecture}
In this Section we revise the ideas behind a generic physics-informed neural network. It is then detailed the particular kind of PINN used here for solving Schr\"odinger's equation, the A-PINN \cite{YUAN2022111260}: Physics-informed neural network with Auxiliary output.
Physics-informed neural networks \cite{RAISSI2019686}, or PINNs are a category of neural networks that aim to solve Partial Differential Equations by encoding prior knowledge about the physical system, most importantly the differential equation itself, in their loss function. This loss can be written as:
\begin{equation} \label{eq: loss_generic}
    f = Loss_{data} + Loss_{PDE} +  Loss_{phys}
\end{equation}
\(Loss_{data}\) is the standard loss for a neural network: take a set of $N$ training points $\{\vec{x_i}\}$ with $0 \le i \le N-1$ and the corresponding labels $\{\vec{y_i}\}$, where $\vec{y_i}$ is the value that the objective function assumes in $\vec{x_i}$. Given the network outputs $\{\vec{out}_{i}\}$ for the inputs $\{\vec{x_i}\}$, \(Loss_{data}\) will be:
\begin{equation}
    Loss_{data} = Err(\{\vec{y}_i\}, \{\vec{out}_{i}\})
\end{equation}
Where Err represents a chosen metric. We will now list the relevant ones for this tutorial:
\begin{equation}
\label{eq: types_of_error}
\begin{aligned}
    \textrm{Mean Squared Error:}& \;\;\;\; MSE(\{\vec{y_i}\}, \{\vec{out}_{i}\}) = \frac{1}{N}\sum_i\|\vec{y_i} - \vec{out}_{i}\|^2 \\
    \textrm{Sum of Squared Errors:}& \;\;\;\; SSE(\{\vec{y_i}i\}, \{\vec{out}_{i}\}) = \sum_i\|\vec{y_i} - \vec{out}_{i}\|^2 \\
    \textrm{Mean Absolute Error:}& \;\;\;\; MAE(\{\vec{y_i}\}, \{\vec{out}_{i}\}) = \frac{1}{N}\sum_i\|\vec{y_i} - \vec{out}_{i}\| \\
    \textrm{Sub of Absolute Errors:}& \;\;\;\; SAE(\{\vec{y_i}\}, \{\vec{out}_{i}\}) = \sum_i\|\vec{y_i} - \vec{out}_{i}\| \\
\end{aligned}
\end{equation}
Note that in our case this \(Loss_{data}\) will not be present since we are working in an unsupervised setting. Those metrics are still used to compute the unsupervised losses. \(Loss_{phys}\) corresponds to the losses given by the physical constraints of the system, like boundary conditions and initial conditions.
Lastly, \(Loss_{PDE}\) is the error for the differential equation. For instance, given a PDE with implicit solution \(\mu(t)\) 
that depends on \(\mu(t)\) and its derivative \(\partial_t\mu(t)\) with respect to the independent variable $t$, the differential equation can be written as:
\begin{equation} \label{eq: generic_PDE}
    \partial_t\mu(t) + N[\mu;\boldsymbol{\lambda}] = 0
\end{equation}
where $N$ is a nonlinear operator and \(\boldsymbol{\lambda}\) is some set of parameters. Given this, we can write \(Loss_{PDE}\) (using MSE as metric) for a network designed to solve this equation as:
\begin{equation} \label{eq: generic_PDE_loss}
    \frac{1}{n}\sum_{i=1}^n(\partial_t\mu(t_i) + N[\mu(t_i);\boldsymbol{\lambda}])^2
\end{equation}
Where \(\{t_i\}\) is the set of $n$ points, called collocation points, where the PDE will be evaluated.
This makes it possible to train neural networks even in the presence of noisy or very little data, and even in a fully unsupervised setting, such as in this paper. 

However, when working with quantum mechanical systems, we are usually not just dealing with Partial Differential Equations, but often with Integro-Differential Equations. At the very least there is always the need to compute an integral to enforce the wave function's normalization. Normally, to compute an integral numerically finite difference methods are employed. Those methods utilise a mesh, a set of finite elements used to represent the function, to compute the integral. However, being mesh-free is one of the advantages of utilizing a neural method. This advantage can be retained if we are able to employ a mesh-free way to compute the required integrals. To this end we employ so-called auxiliary outputs: beyond the wave function, we add an additional output to the network for each integral we need to compute. For instance, an integral is needed for the normalization in a domain \([-\frac{L}{2}, \frac{L}{2}]\). It can be represented by the auxiliary output \(\nu(x)\), defined as:
\begin{equation}
\label{eq: integral}
    \nu(x) = \int_{-\frac{L}{2}}^x |\psi(k)|^2\, dk\
\end{equation}

and the normalization will be fulfilled if \(\nu(\frac{L}{2}) = 1\); \(\nu(-\frac{L}{2}) = 0\). Note that there is a trade-off here. On the one hand, the discretization error can be potentially removed using an auxiliary output. On the other hand, using it makes the network harder to train due to the additional outputs and losses. Therefore the choice to utilize an Auxiliary output or a finite difference method has to be made on a case-by-case basis.

\subsection{Potential analysis} \label{sec: preliminaries}
This Section enters into the specifics of how to build a PINN to solve Schr\"odinger's equation. 
Here we focus in particular on the analysis of the potential to find the optimal representation of the problem.
The potential is then used to define the associated Schr\"odinger equation. Beyond that, it is also important to identify the independent variables that will make up the input of the neural network. Likewise, we must identify any integrals that will make up the auxiliary output. For instance, for an infinite potential well, the representation of the Schr\"odinger's equation will just contain the kinetic part, and the potential:
\begin{equation}
    \label{eq: potential_well}
    \begin{cases}
      V(x) = 0 \;\;\;\;for\,x\in[-\frac{L}{2}, \frac{L}{2}]\\
      V(x) = \infty \;\;\;\; otherwise
    \end{cases}\,.
\end{equation}
can be represented by having \([-\frac{L}{2}, \frac{L}{2}]\) be the computational domain and imposing Dirichlet boundary conditions:
\begin{equation}\label{eq: bc}
    \psi(-\frac{L}{2}) = \psi(\frac{L}{2}) = 0
\end{equation}
The position $x$ will be the input. The outputs will be the main one, $\psi(x)$, and an auxiliary one, $\nu(x)$, as defined in \eqref{eq: integral}.\\

\subsection{Physical constraints and inductive biases}  \label{sec: constraint}
After defining the main differential equation loss in the previous Section, here we explain the other losses used to build a neural network capable of solving the equation. In particular, here are defined the losses linked to the physical constraints of the system. We also add a kind of loss not essential to solving the equation but which greatly improves convergence, namely the inductive biases. 

To define these losses one needs to understand which are the physical constraints of the system and if any symmetries can be exploited to obtain an inductive bias. By inductive bias, we mean a set of conditions that leads the network to prioritize a set of solutions over another, for instance prioritizing solutions that follow a given symmetry. Biases can be invaluable because they limit the search space, and thus reduce the number of epochs needed to obtain a solution.\\
A condition that is always needed is the normalization:
\begin{equation}
\label{eq: normalization_gen}
    \int_{\Omega} |\psi(\vec{x})|^2 \,d\vec{x} \, = 1
\end{equation}
 where $\Omega$ is the domain of the wave function, chosen depending on the dimensionality and nature of the system. 
 For instance, for a one-dimensional infinite potential well in \([-\frac{L}{2}, \frac{L}{2}]\), $\Omega$ can be restricted to \(\Omega = [-\frac{L}{2}, \frac{L}{2}]\). It is important to note that imposing the condition defined in \eqref{eq: normalization_gen} is the only thing that forces the wave function to be nonzero. Otherwise, $\psi(x) = 0$ for all $x$ would exactly satisfy all losses. This integral can be solved by utilizing an auxiliary output $\nu(x)$ with form \eqref{eq: integral} and by imposing \(\nu(\frac{L}{2}) = 1\); \(\nu(-\frac{L}{2}) = 0\). However, we must first train the network so that $\nu(x)$ has the desired form \eqref{eq: integral}. Once again, this can be achieved by exploiting automatic differentiation, to compute $\grad\nu(\vec{x})$ and thus use the following as a loss:
\begin{equation} \label{eq: integ_loss}
    \grad\nu(\vec{x}) = |\psi(\vec{x})|^2
\end{equation}
obtaining the integral loss.
We typically need to set some boundary conditions. In general, one can either use hard or soft boundary conditions. The former choice forces the network output to fulfill the boundary conditions exactly. This can be done by employing a change of variable or modifying in some way the output, such as by multiplying it by another function. The latter choice makes the network learn the BCs via an additional loss, the boundary condition loss. This loss is $0$ when the BCs are fulfilled. For instance in the infinite potential well to impose soft Dirichlet Boundary conditions \eqref{eq: bc} we can add to the network's loss $|\psi(-\frac{L}{2})| + |\psi(\frac{L}{2})|$. 
In this tutorial, we always use soft boundary conditions. 

Further constraints can be required on a case-by-case basis, but normalization and boundary conditions are always essential. Moreover, we can also employ an additional kind of loss: the inductive biases. These are losses that are not required to obtain a solution, but can instead be added to allow the network to converge much faster.
They are usually based on the symmetries of the potential: for instance, the infinite well is symmetric with respect to $0$. This symmetry makes its eigenfunctions alternatively even and odd with respect to $0$. A conditions that can be enforced in the PINN by the symmetry loss:
\begin{equation} \label{eq: symm_loss}
    \psi(x) - s\psi(-x) = 0
\end{equation}
where $s$ is $1$ for symmetric states and $-1$ for antysimmetric ones (the value of $s$ is switched each time a network is trained, starting from $s$=1). \cor{In general, we can utilize as inductive biases any kind of additional constraint that we know the wave function must follow. The implementation depends on the inductive bias we want to add, but in general it can be added as an additional loss. Like all neural network losses it must always be nonnegative, and must reach $0$ when the bias is satisfied. It is usually effective to give such additional losses an higher weight compared to the other losses, in order to let the inductive bias better fulfill its role as a condition that limits the space of the solution. For instance in the example in Eq. \eqref{eq: symm_loss} the total loss will greatly increase for each pair of points that does not follow the symmetry constraint. This leads the neural network to only explore the space of symmetric or anti-symmetric solutions (depending on the value of $s$).}

\subsection{Self-consistent computation of the eigenenergy}  \label{sec: energy}
In this Section, we explain how to find the eigenvalue to pair with the eigenfunction obtained thanks to the techniques described in the previous Sections.

Currently, the network can only find the eigenfunction of the potential when the energy is known. However, in general when solving the Schr\"odinger equation that will not be the case. Therefore a way to find it in parallel with the eigenfunction is needed. There are two ways to do this: the first is to set the energy $E$ itself as a trainable parameter, but this seems to result in fairly slow training. The second is to build a second neural network to train in parallel with the one that predicts the wave function. It will take as input a constant stream of $"1"$s of the same length as the input vector, and output a guess for the energy $E$. Since the output is constant, this second network will have no hidden layers and there will be no nonlinear activation function. The network will just perform a linear transformation from $1$ to the value of the energy, based on a trainable parameter.

\subsection{Finding the desired state} \label{sec: orthogonality}
A crucial step is to build the losses that allow the network to distinguish the desired eigenstate from the infinite spectrum of the Hamiltonian. This is explained in this Section.
By training the PINN with the current losses, it will solve the Schr\"odinger equation for a random state anywhere in the spectrum. Usually, however, we are looking for some amount of eigenstate starting from the ground state and going in order of increasing energy. In particular, it is especially important to build a network that finds the ground state first. This can be achieved by introducing an energy minimization condition in the loss:
    \begin{equation}\label{eq: loss_en}
    e^{a(E_{PINN} - E_{init})} = 0
    \end{equation}
where \(E_{PINN}\) is the predicted energy, $a$ is a hyperparameter. \(E_{init}\) is either another hyperparameter when training the network to predict the ground state, or the energy of the previous state, for the excited states. The choice to increase \(E_{init}\) as the quantum number increases has been made to avoid the loss exploding for higher energy states due to its exponential increase. Otherwise, the network might fail to converge at all. We also have to take into account that this energy minimization loss keeps pushing the network to lower energies even when the correct solution is reached. This moves the global minima of the loss from the eigenstate to a lower energy state, causing the loss to converge to an incorrect state. 
To avoid such a situation, the weight of the loss is gradually decreased as the training continues. This reduction should be such that the loss becomes approximately $0$ towards the end of the training. Thanks to this in the earlier epochs this loss will steer the network towards the desired lower energy state. On the other hand in the latter epochs, the energy minimization loss will become negligible and allow the network to settle to the correct solution. The resulting neural network can now find the ground state of the wave function. We now want to obtain the excited states. To do so, a sequence of networks is built, each of which predicts one of the Hamiltonian's eigenstates and the corresponding eigenvalue, in order of increasing energy. To do so, an additional loss is employed, the orthogonality loss. This loss forces each network's output to be orthogonal to all the ones before it:
    \begin{equation} \label{eq: orth_loss}
        \sum_i\braket{\psi(\vec{x})}{\psi_i(\vec{x})}{} = 0
    \end{equation}
where \(\psi_i(\vec{x})\) are the known eigenfunctions. This loss will, of course, always be $0$ when training the first network. But each time a network converges it will be saved and utilized to obtain the \(\psi_i(\vec{x})\) needed to compute this loss for the networks following it in the sequence.

\subsection{Weights} \label{sec: weights}
In the previous Sections, the losses to minimize have been defined. In the following Sections is explained how to build the actual computational framework that ensures the network is able to minimize those losses. In this Section, in particular, some pointers are given on how to tune one of the most important set of hyperparameters: the weights.
Usually, when training PINNs it is better to set weights that ensure all the losses have the same scale. That is not the case for this problem: it is very easy for a network that has to predict eigenfunctions to get stuck in local minima such as $\psi(\vec{x}) = 0$ $\forall \vec{x}$ or an already discovered eigenvector. This means that it is actually better to set the weights in such a way that the network first enforces the physical constraints of the system, such as normalization and orthogonality. Only once these are fulfilled the network should be trained to actually solve the equation. This means setting the losses for those physical constraints higher than those of the differential equations. 

In general, this means that the integral loss should have the highest weight, then the normalization, needed to avoid the constant solution, and the orthogonality, needed to avoid repeating solutions. At this point, the network should minimize the boundary conditions loss. The last loss to minimize is the differential equation loss. However, since solving the differential equation is the main objective of the network it is often worth it to gradually increase its loss' weight. This allows the network to learn how to enforce the physical conditions in the first epochs but makes it focus more on the differential equation in the latter epochs. It is important to stress that weights are probably the most impactful hyperparameter to tune: a correct choice of weights will be the difference between quick convergence and a network getting stuck on a suboptimal solution.

\subsection{Choice of training points} \label{sec: training points}
In order to train the PINN, the losses must be evaluated on a set of points, known as collocation points. These can be taken anywhere in the domain, and this Section contains the optimal way to choose them.
When training a PINN such as this, each batch of training points should span the whole domain. Furthermore, since no dataset is being used, the choice of points is not constrained in any way: they can be taken anywhere in the domain. This can be exploited to basically work with an infinitely large dataset. To fully take advantage of this, first a batch size $N$ is decided. Then a mesh of $N$ points that spans the whole domain is built. Each of these points will be the center of a normal distribution from which one of the points in the given batch will be sampled. This results in batches that are always slightly different but always span the whole domain, thus ensuring the maximum exploration and fastest possible convergence.

\subsection{Domain definition} \label{sec: domain}
This computational framework inevitably requires the collocation points to be chosen within a certain domain. In this Section, the best way to identify this domain is explained.
The choice of the computational domain is crucial for an effective convergence of the network. First of all, this method only makes sense if almost all of the nonzero part of the desired wave function exists within the chosen domain. This means that outside it the wave function must be either exactly $0$ or vanishingly small and thus negligible. On the other hand, if the computational domain is too wide then the network will be trained on a vast area of just $0s$ that will provide no information to it, resulting in a greatly slowed down training. This means that more points in each batch would be needed to properly sample the "interesting" part of the domain. In the case of the well the choice of the optimal domain is an immediate consequence of the structure of the system: as mentioned in Section \ref{sec: preliminaries} just using \([-\frac{L}{2}, \frac{L}{2}]\) is ideal since the wave function will be nonzero only inside the well. For more complex potentials the choice of the domain must be made based on an analysis of the potential under study. One needs to find out at which distance from the origin the wave function becomes negligible. Doing this will probably require some trial and error. 

\subsection{Network architecture} \label{sec: NN}
In this Section, we give some suggestions and tricks on how to better build a feed-forward neural network to solve the Schr\"odinger in a physics-informed way. We focus on which metrics to use for the loss and which activation function should be employed in the network.
PINNs tend to work better as relatively shallow but wide networks. For instance, in \cite{brevi2024addressing} the main network was made up of $7$ hidden layers of $256$ neurons each. \cor{In general, the physics-informed approach can can be applied to any kind of neural network, since its implementation is based on modifying just the loss function, independently of the network's architecture. Here we choose to focus on feed-forward PINNs due to it being the simplest architecture, which makes it ideal for a tutorial. The number of hidden layers, of course, depends on the problem we need the PINN to address. For the examples given in this tutorial, we employed a PINN with 6 hidden layers to solve the infinite potential well and one with 10 hidden layers to find the spectrum of the particle in a ring.} As for the activation function, an infinitely differentiable one such as $tanh$ is needed. Otherwise, the network would be unable to compute higher-order derivatives via automatic differentiation. Choosing an activation function of this type instead of something like $ReLU$ might, however, cause vanishing gradient problems. If this happens, it might be worth it to look into adaptive activation functions \cite{adaptive_activation_function} as a solution. \cor{For this tutorial, we always employ $tanh$ as the activation function. The outputs of each layer was not normalized, since empirically it was found that output normalization does not improve convergence.} Lastly, it is not necessary to compute all the partial losses using the same metric. It usually makes more sense to utilize something like the Sum of Absolute Errors for the losses enforcing the physical conditions of the system, such as the Boundary Conditions loss or the Normalization loss. This is because the rest of the training makes sense only if those losses are approximately $0$ in all points. And this is enforced more strictly with SAE rather than something like MSE. On the other hand, Mean Squared Error is usually the best choice for the integral and differential equation losses. Utilizing the Sum of Squared Error to compute these losses is also worth considering. This is especially true when the wave function is expected to have a complex behavior in a small section of the domain and a relatively simple one in the rest. This choice of metric will avoid underestimating the loss due to the averaging of the error for the points in the large "simple" area and the small "complex" area. As for the optimizer, Adam \cite{KingBa15} is usually rather effective. 

\section{How to evaluate training} \label{sec: evaluation}
In this section, some ways to evaluate the PINN's training to optimize their architecture and hyperparameters are shown. The focus will be on some typical pathological behaviors and some ways to understand what causes them and how to fix them.
The behavior of the loss actually tends to be a pretty reliable metric when training PINNs in an unsupervised way, and is actually more reliable than for standard neural networks. This is because a PINN follows a given physical law, and the loss quantifies how well its output follows the given law. It does not have a set of training points, it might memorize instead of learning the underlying law. 

As such, overfitting on some training sets is not a risk and it does not make sense to build the training, the validation and the test sets from a dataset. In spite of this, due to the complexity of the loss function just looking at the full loss does not tell the whole story of what is happening during training. In the following subsections, some ways to interpret the behavior of the network during training more deeply will be detailed. This should allow one to make more informed decisions when tuning hyperparameters.

\subsection{Partial losses} \label{sec: partial_losses}
The first thing to look at when the PINN does not converge is to try to understand which loss or losses is failing to converge. At this point, the first thing one can do to try and improve the training can be to try to either increase the weight for the losses that are struggling to converge or reduce the weights for other, competing losses. One very common problem is the orthogonality loss failing to converge while other losses get very close 0. This means that the network is predicting an already discovered eigenfunction \footnote{In low dimensions this can also be seen by looking at partial plots}. For this increasing the weight of the orthogonality loss is usually a pretty safe solution.\\
Another common occurrence is the normalization loss staying high while all other losses become 0. This usually means that the network has converged to the local minima $\psi(\vec{x}) = 0$ $\forall \vec{x}$. While it could be fixed by increasing the normalization loss' weight \footnote{this usually means also increasing the normalization integral's weight, since the normalization loss does not have any meaning if the normalization integral is incorrect} it can also be useful to change the initialization of the network, for instance going from Xavier uniform \cite{xavier_init} to Kaiming normal \cite{kaiming_he_init}, in order to start the network further from this local minima. This could also be a sign of the energy minimization loss having excessive weight or growing too quickly. A way to decide this in low dimensions is to display the plot of the best wave function (i.e., the one with the lowest loss) every number of epochs, and look for any pathological behaviour in its evolution. Another example is the case in which the boundary condition loss seems to fail to converge: in this case if the plotted wave function resembles what we expect for the given potential but does not vanish at the border of the domain it might mean that a larger domain is needed. It is important to note that a plateauing loss is not always a sign of a failed convergence. While training it might happen that the PINN's loss remains stuck around a certain value for thousands of epochs but will start decreasing again after some time. On the other hand, wild oscillations of the loss might be a sign of an excessively high learning rate. Note that these are only some empirical heuristics and might not apply in all cases. It is important to look at the loss' behaviour in a case-by-case basis and adjust the weights depending on what is best in the current circumstances.

\subsection{Evaluation metrics} \label{sec: metrics}
In the previous section only the losses have been utilized to understand how successful the training has been. However, depending on the problem other, more informative, metrics might be available. In this section two metrics that allow to gauge the goodness of PINN's solution for the Schr\"odinger equation are detailed. 
When a solution is already known due to different numerical methods or because the potential has an analytic solution, the correctness of the PINN's output can be easily evaluated by utilizing fidelity \cite{fid1, fid2, fid3} and relative error on the energy. To gauge the eigenvalue's correctness the relative error for the energy is used:
\begin{equation}\label{eq: rel_error_e}
    err_E = \frac{E_{Ex} - E_{PINN}}{E_{Ex}}
\end{equation}
where \(E_{Ex}\) is the analytic value of the energy, while \(E_{PINN}\) is the one predicted by the network.
On the other hand to evaluate the eigenvector the fidelity between the predicted wave function and the exact one is employed:
\begin{equation}\label{eq: sim_wf}
    \mathcal{F}_\psi = |\braket{\psi_{Ex}}{\psi_{PINN}}{}|^2
\end{equation}
 numerically evaluated as:
 \begin{equation}\label{eq: sim_wf, discrete}
     \mathcal{F}_\psi = |\sum_i \psi_{PINN}(\vec{x}_i)\overline{\psi_{Ex}(\vec{x}_i)}|^2
 \end{equation}
where the {\(x_i\)} are a set of points spanning the whole domain, and \(\psi_{Ex}(x)\) and \(\psi_{PINN}(x)\) are the values of the wave function at the point $x$ as given by the analytic expression and by the network, respectively. Both of the vectors will be normalized before calculating the similarity, to avoid getting incorrect results due to inaccurate scaling.
This should equal one when the two wave functions are the same (or if their difference is just a nonphysical phase).




\section{Examples} \label{sec: examples}
Some examples of quantum mechanical systems that can be solved by using PINNs are now shown: the Infinite potential well and the particle in a ring.

\subsection{Infinite potential well} \label{sec: well}
In this section we study the infinite potential well. This system has been chosen due to its extreme simplicity, making it an excellent testbed to build the basis of the method.
We are studying, in particular, the infinite potential well \([-\frac{L}{2}, \frac{L}{2}]\), with $L=3$:
\begin{equation}
    \label{eq: potential_well_example}
    \begin{cases}
      V(x) = 0 \;\;\;\;for\,x\in[-\frac{3}{2}, \frac{3}{2}]\\
      V(x) = \infty \;\;\;\; otherwise
    \end{cases}\,.
\end{equation}
As mentioned in section \ref{sec: preliminaries} this potential can be represented just by studying the region with $V(x) = 0$ with Dirichlet boundary conditions. Therefore, based on the previous sections, the neural network takes as input the position $x$ and outputs the wave function $\psi(x)$ and auxiliary output \(\nu(x)\). The network's losses will be:
\begin{equation}
\label{eq: Losses_well}
\begin{aligned}
    \textrm{Integral loss:}& \;\;\;\; \left. \frac{\partial \nu(x')}{\partial x'}\right|_{x' = x} = |\psi(x)|^2 \\
    \textrm{Normalization loss:}& \;\;\;\; \nu(-\frac{L}{2}) = 0, \;\; \nu(\frac{L}{2}) = 1 \\
    \textrm{Boundary conditions loss:}& \;\;\;\; \psi(-\frac{L}{2}) = \psi(\frac{L}{2}) = 0 \\
    \textrm{Energy minimization loss:}& \;\;\;\; e^{a(E_{PINN} - E_{init})} = 0,\;\; with\;\; a = 0.8;\;\; E_{init} = 0\;\; when\;\; n = 0 \\
    \textrm{Orthogonality loss:}& \;\;\;\; \sum_i\braket{\psi(x)}{\psi_i(x)}{} = 0 \\
    \textrm{Differential equation loss:}& \;\;\;\; -\frac{1}{2}\frac{\partial^2 \psi(x)}{\partial x^2} - E \psi(x) = 0 \\
\end{aligned}
\end{equation}



To those losses an inductive bias can be added, since the potential is symmetric with respect to 0. Due to this symmetry, The wave functions that solve the equation are alternatively symmetric and anti-symmetric with respect to 0. This condition can be enforced by adding the loss:
    \begin{equation}
    \label{eq: symm-loss}
    \psi(x) - s\psi(-x) = 0    
    \end{equation}
where $s$ is a parameter that assumes the value one if we are looking for a symmetric state ($n$ even), 
or $s=-1$ if we are looking for an anti-symmetric one ($n$ odd). For the sake of clarity, it is worth mentioning that 
the value of $s$ starts at one and is switched every time a new model is trained.\\
The weight for each of these losses are summarized in table \ref{tab: Well_weights}. Note how they follow the heuristics in Section \ref{sec: weights}.

\begin{table}[H] 
\caption{Weights for the infinite potential well.\label{tab: Well_weights}}
\begin{tabularx}{\textwidth}{CCC}
\toprule
\textbf{Loss}	& \textbf{Weight}\\
\midrule
Normalization & 1000			\\
Boundary conditions & 10			\\
Integral & 5000			\\
Orthogonality & 1000			\\
Symmetry & 1000 \\
Energy Minimization (starting) & 10			\\
Differential equation (starting)		& 1			\\
\bottomrule
\end{tabularx}
\end{table}

6 networks are trained, predicting the ground state and first $5$ excited states. Each of them has $6$ hidden layers of $64$ neurons, and is trained on batches of $512$ collocation points. The hyperbolic tangent has been used as the activation function.

All networks quickly converge\footnote{The network is considered converged if the total loss is less then $10^{-1}$ and the differential equation loss is less then $5\times10^{-3}$. These values can be reduced to obtain greater accuracy at the cost of a longer training run. Note that a patience condition cannot be utilized due to the tendency of the PINN's loss to plateau for a long period before it resumes going down.}. The exact eigenfunctions and eigenvalues of the well are known analytically:

\begin{equation}
\label{eq: well_functions}
    \begin{aligned}
        \psi(x) =& \sqrt{\frac{2}{L}} sin\left( n\frac{\pi}{L} \left( x + \frac{L}{2} \right)  \right) \\
        E_n =& \frac{n^2\pi^2}{2L^2}
    \end{aligned}
\end{equation}

Therefore, we can use the metrics in Section \ref{sec: metrics} to judge the performance of the network. The results are reported in table \ref{tab: Well_res}.

\begin{table}[H] 
\caption{Metrics for the infinite potential well.\label{tab: Well_res}}
\begin{tabularx}{\textwidth}{CCC}
\toprule
\textbf{Quantum number n}	& \textbf{$err_E$}	& \textbf{$\mathcal{F}_\psi$}\\
\midrule
1 (Ground state)		& $2.68\times10^{-4}$			& $0.9999943$\\
2		& $2.68\times10^{-4}$			& $0.9999943$\\
3		& $-6.07\times10^{-4}$			& $0.9999910$\\
4		& $-7.65\times10^{-4}$			& $0.9999955$\\
5		& $-9.82\times10^{-5}$			& $0.9999910$\\
6		& $1.78\times10^{-5}$			& $0.9999998$\\
\bottomrule
\end{tabularx}
\end{table}

\begin{figure*}[htbp]
    \centering
    \begin{subfigure}[b]{0.31\columnwidth}
        \includegraphics[width=\linewidth]{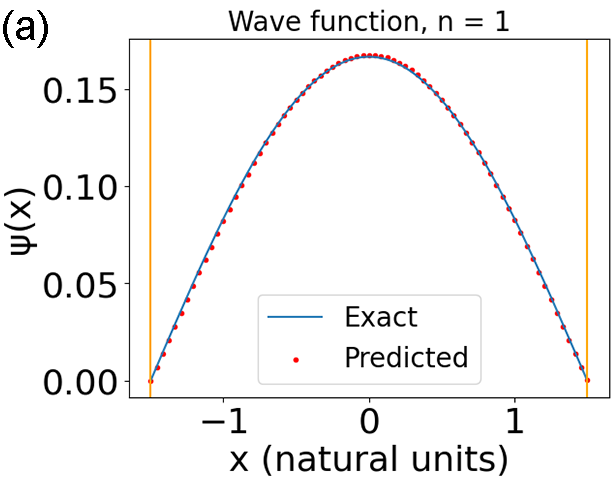}
    \end{subfigure}
    \begin{subfigure}[b]{0.31\columnwidth}
        \includegraphics[width=\linewidth]{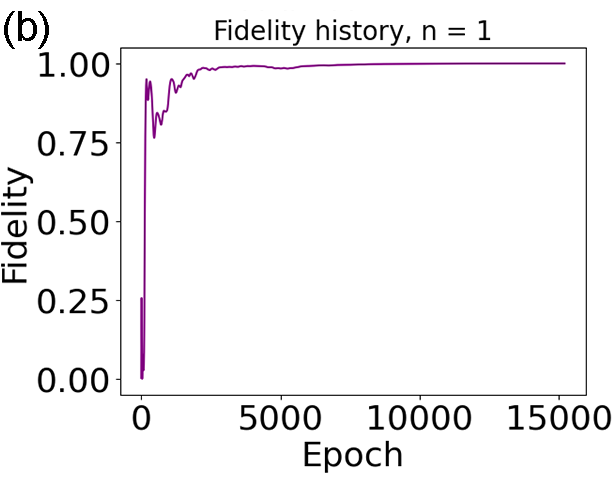}
    \end{subfigure}
    \begin{subfigure}[b]{0.31\columnwidth}
        \includegraphics[width=\linewidth]{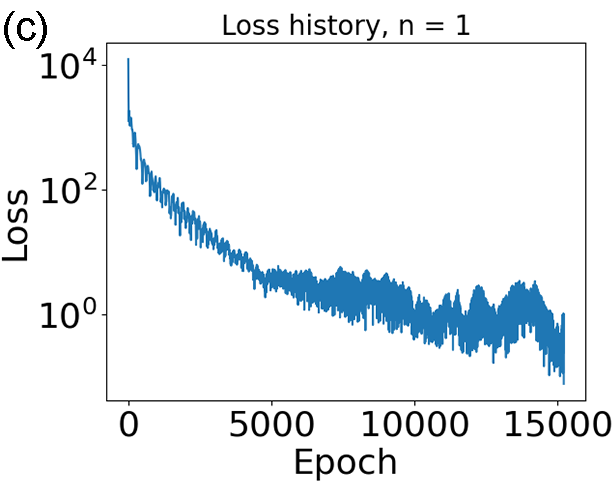}
    \end{subfigure}
    \caption{(a) Ground state for the well. The red dots are the PINN's predictions, while the blue line is the ground truth. The yellow vertical lines represent the walls of the well. (b) Fidelity throughout training for the ground state. (c) Loss behavior throughout training. The y axis is in logarithmic scale, therefore the oscillations on the y axis for low losses are overemphasized }
    \label{fig: plots_well_n1}
\end{figure*}

\begin{figure*}[htbp]
    \centering
    \begin{subfigure}[b]{0.31\columnwidth}
        \includegraphics[width=\linewidth]{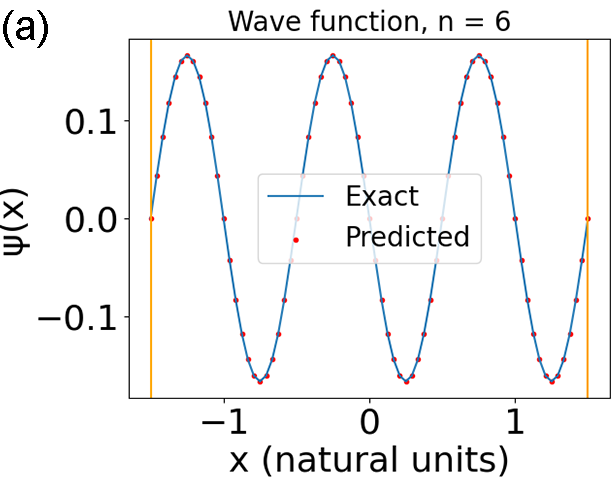}
    \end{subfigure}
    \begin{subfigure}[b]{0.31\columnwidth}
        \includegraphics[width=\linewidth]{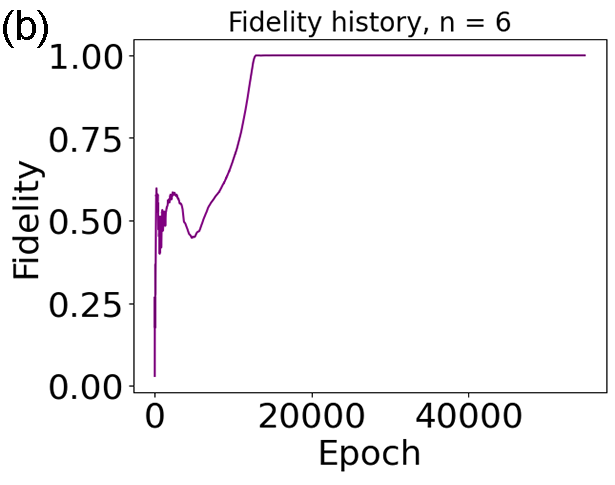}
    \end{subfigure}
    \begin{subfigure}[b]{0.31\columnwidth}
        \includegraphics[width=\linewidth]{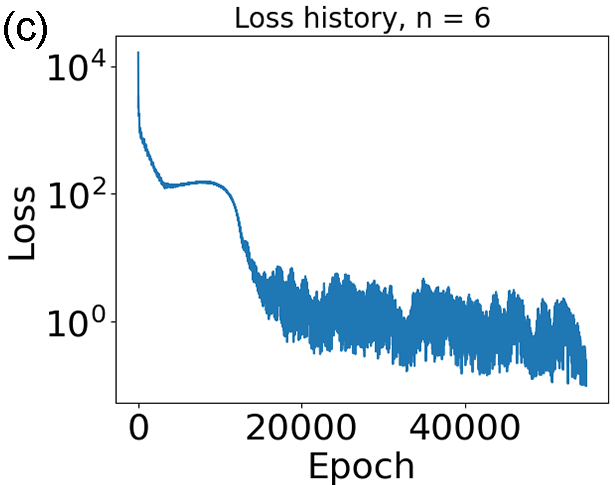}
    \end{subfigure}
    \caption{(a) Fifth excited state for the well Eq.\eqref{eq: well_functions}. The red dots are the PINN's predictions, while the blue line is the ground truth. The yellow vertical lines represent the walls of the well. (b) Fidelity throughout training for the ground state. (c) Loss behavior throughout training. The y axis is in logarithmic scale, therefore the oscillations on the y axis for low losses are overemphasized}
    \label{fig: plots_well_n6}
\end{figure*}

It is now useful to look at the final plots for the ground state and the last excited state. Starting with the ground state (Figure \ref{fig: plots_well_n1}) one can immediately see from Figure \ref{fig: plots_well_n1}a that the neural network prediction closely aligns with the ground truth. Furthermore, by looking at the evolution of the fidelity in Figure \ref{fig: plots_well_n1}b it can be seen that it quickly reaches a decent value but after the first few thousand epochs and then starts growing very slowly until it approaches 1
This behavior of the fidelity is also matched by the behavior of the loss (Figure \ref{fig: plots_well_n1}c), decreasing monotonically but more slowly in the latter epochs.
For the fifth excited state (Figure \ref{fig: plots_well_n6}), the situation is slightly different. In particular, the wave function itself in Figure \ref{fig: plots_well_n6}a shows good convergence, given a qualitative look. It is interesting to look at fidelity in Figure \ref{fig: plots_well_n6}b. This graph once again  grows very quickly in the first epochs, then the increase in fidelity stagnates until it goes back to increasing until it saturates. This can actually already be seen in \ref{fig: plots_well_n1}b but is much more marked here. This behavior is once again matched by the loss in Figure \ref{fig: plots_well_n6}c, which plateaus around the same time as the fidelity decreases, only to resume decreasing at decent speed after a couple thousand epochs. This is a kind of pathological behaviour that is very common for PINNs, and is probably further exacerbated by the presence of the energy minimization condition. This condition may be causing the network to not converge early since the energy minimization loss would be too high. The network converges only once that loss' weight becomes approximately $0$

\subsection{Particle in a ring}
In this Section we show another example of potential: the particle in a ring. In this case, the potential is $0$ on a ring of radius L, and $\infty$ everywhere else. Using polar coordinates, the wave function depends only on the angular coordinate $\theta$. The shape of the potential actually becomes the same as the infinite potential well of Section \ref{sec: well}, with the difference that instead of having Dirichlet boundary conditions it requires periodic boundary conditions, that is:

\begin{equation}
\label{eq: periodic_bc}
    \psi(0) = \psi(2\pi)
\end{equation}
Or, more in general, we require:
\begin{equation}
\label{eq: periodicity}
    \psi(\theta) = \psi(\theta + 2\pi)
\end{equation}

Furthermore, the change of variable makes the gradient assume the following form:
\begin{equation}
    \grad^2 = \frac{1}{L^2}\frac{\partial^2}{\partial \theta^2}
\end{equation}

This makes the analytical form of the solutions:

\begin{equation}
\label{eq: ring_functions}
    \begin{aligned}
        \psi(\theta) =& \frac{1}{\sqrt{2\pi}}e^{in\theta} \\
        E_n =& \frac{n^2}{2L^2} \;\;\; where\; n=0,\pm 1, \pm 2, \pm 3, \dots
    \end{aligned}
\end{equation}

This actually introduces two significant challenges in training the PINN: first of all, $\psi(\theta)$ will in general be a complex number, which means that the main output of the network will now be two dimensional, where the two outputs are $Re\left(\psi(\theta)\right)$ and $Im\left(\psi(\theta)\right)$.  Furthermore, each of the eigenstates of the ring with the same absolute value of $n$ is degenerate. This means that the neural network will, in general, find an arbitrary linear combination of the two degenerate states. This still results in correct solutions, but will make calculating metrics such as fidelity more complex. Therefore, in order to make the results more readable for this tutorial we impose an additional loss that forces the two wave functions to have the same norm. This allows the solution to converge to \eqref{eq: ring_functions}, instead of any linear combination of degenerate states. The network's losses will be:
\begin{equation}
\label{eq: Losses_ring}
\begin{aligned}
    \textrm{Integral loss:}& \;\;\;\; \left. \frac{\partial \nu(\theta')}{\partial \theta'}\right|_{\theta' = \theta} = |\psi(\theta)|^2 \\
    \textrm{Normalization loss:}& \;\;\;\; \nu(0) = 0, \;\; \nu(2\pi) = 1 \\
    \textrm{Boundary conditions loss:}& \;\;\;\; \psi(0) = \psi(2\pi) \\
    \textrm{Periodicity loss:}& \;\;\;\; \psi(\theta) = \psi(\theta + 2\pi) \\
    \textrm{Energy minimization loss:}& \;\;\;\; e^{a(E_{PINN} - E_{init})} = 0,\;\; with\;\; a = 0.4;\;\; E_{init} = 0\;\; when\;\; n = 1 \\
    \textrm{Orthogonality loss:}& \;\;\;\; \sum_i\braket{\psi(\theta)}{\psi_i(\theta)}{} = 0 \\
    \textrm{Differential equation loss:}& \;\;\;\; -\frac{1}{2L^2}\frac{\partial^2 \psi(\theta)}{\partial \theta^2} - E \psi(\theta) = 0 \\
\end{aligned}
\end{equation}

Here the Periodicity loss and Boundary conditions loss are treated as two separate losses in order to utilize two different metrics to compute them. The boundary conditions loss is computed with SAE and ensures the correct behavior of the wave function at the domain's boundaries. On the other hand, the periodicity loss is computed with SSE and acts more as an inductive bias. The main neural network is made up of $10$ layers of $256$ neurons each. $1024$ collocation points were employed. 
The results for $L = 0.95$ are reported in table \ref{tab: Ring_res}. Note that for the trivial ground state $\psi(\theta) = \frac{1}{\sqrt{2\pi}}$ we relay the predicted energy instead of the relative error for the energy. This is because for this state $E=0$, and this metric is thus undefined.

\begin{table}[H] 
\caption{Metrics for the infinite potential well.\label{tab: Ring_res}}
\begin{tabularx}{\textwidth}{CCC}
\toprule
\textbf{Quantum number n}	& \textbf{$err_E$}	& \textbf{$\mathcal{F}_\psi$}\\
\midrule
0 (Ground state)		& $2.54\times10^{-3}$			& $0.99992261$\\
+1		& $1.48\times10^{-2}$			& $0.9998474$\\
-1		& $1.22\times10^{-2}$			& $0.9996434$\\
\bottomrule
\end{tabularx}
\end{table}

\begin{figure*}[htbp]
    \centering
    \begin{subfigure}[b]{0.31\columnwidth}
        \includegraphics[width=\linewidth]{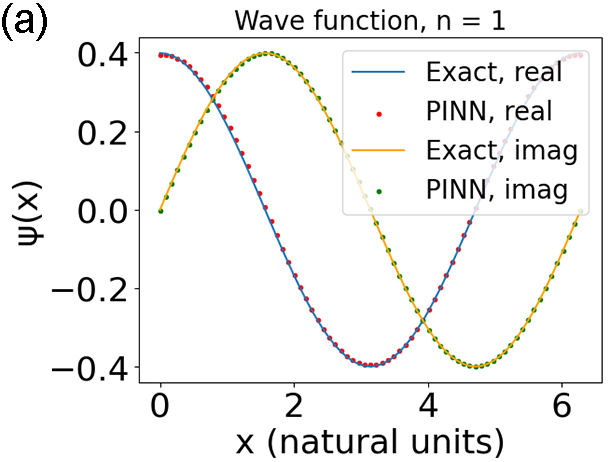}
    \end{subfigure}
    \begin{subfigure}[b]{0.31\columnwidth}
        \includegraphics[width=\linewidth]{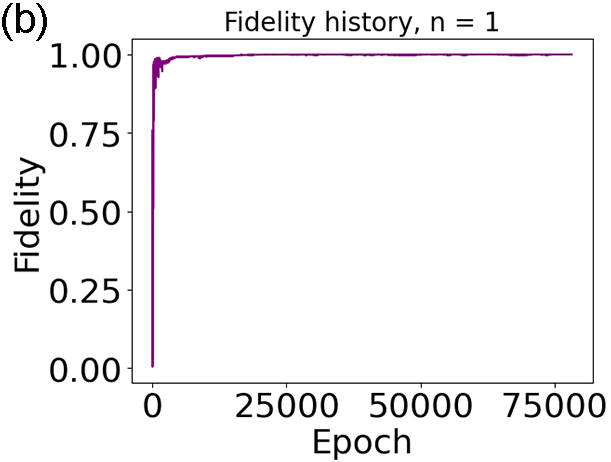}
    \end{subfigure}
    \begin{subfigure}[b]{0.31\columnwidth}
        \includegraphics[width=\linewidth]{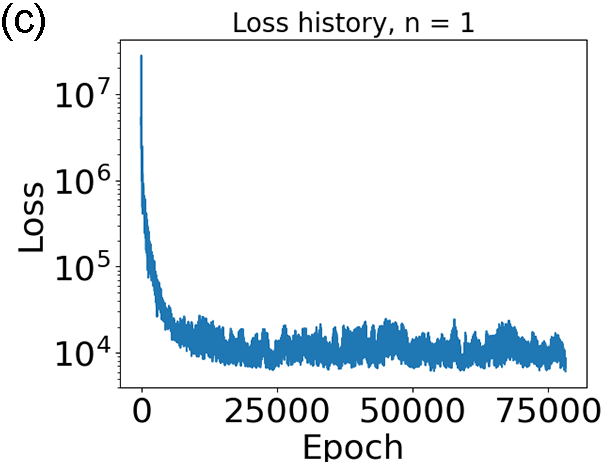}
    \end{subfigure}
    \caption{(a) One of the two first excited states for the ring Eq.\eqref{eq: ring_functions}. The red dots are the PINN's predictions, while the blue line is the ground truth. Note that in this case the ground truth and the neural prediction are actually different states of the pair of degenerate states for $|n|=1$. This discrepancy is due to the fact that the two minima given by the states for $n=1$ and $n=-1$ are equally valid, and thus the neural network can converge freely to either one. However, In order to have fidelity be an informative metrics the real and imaginary parts of the neural network predictions are first multiplied by $sign\left(\braket{Re\left(\psi_{Ex}\right)}{Re\left(\psi_{PINN}\right)}\right)$ and $sign\left(\braket{Im\left(\psi_{Ex}\right)}{Im\left(\psi_{PINN}\right)}\right)$ respectively, resulting in the two plots superimposing in the figure. (b) Fidelity throughout training for the ground state. (c) Loss behavior throughout training. The y axis is in logarithmic scale, therefore the oscillations on the y axis for low losses are overemphasized}
    \label{fig: plots_ring_n1}
\end{figure*}

It is also useful to look at the plot for the state $n=+1$ Figure \ref{fig: plots_ring_n1}. In Figure \ref{fig: plots_ring_n1}a there are the plots of the outputs of the second network compared to the ground state for $n=1$. Due to the fact that $n=1$ and $n=-1$ are degenerate states, however, the network there predicted the state for $n=-1$. This is not an issue since the two are at the same energy, and the only condition asked to the network is that it gives valid eigenstates in order of increasing energy. Furthermore, since the difference between the two are just the sign of either the real or imaginary part, depending on whether or not a global phase is also present, it is possible to modify the plot so that the real and predicted version of the same state are plotted. This is done by multiplying one of the two by the relative phase of the real and imaginary part (also see the caption of Figure \ref{fig: plots_ring_n1}). The result is the plot in Figure \ref{fig: plots_ring_n1}a. Once again the plots for the fidelity, Figure \ref{fig: plots_ring_n1}b, and the loss, Figure \ref{fig: plots_ring_n1}c, have roughly the same behavior. In this case there is no elbow, however they only really improve around the first $15000$ epochs, and only reach the convergence condition very late. Note that here the values of the loss are much larger then in the previous case. This is because SSE was employed in place of MSE, since empirically it can be seen that it leads to better convergence for this system.

\section{Challenges, limitations and poential of PINNs}
 \cor{PINNs have the potential to be able to scale rather well with the complexity of the system. This is thanks to the extreme expressive power of Deep Neural Networks, which are known to be able to successfully approximate any continous function \cite{Universal-approx1, Universal-approx2, Universal-approx3, Universal-approx4}, given a sufficient amount of parameters. However, PINNs do have limitation and disadvantages compared to other computational methods. First of all, the multiple losses create an extremely complex loss landscape, which can be very frustrated \cite{failure-modes}. This makes training the PINN rather difficult, and makes it very likely for it to get stuck in a local minima if the hyperparameters have not been set correctly or if the initialization was especially unlucky. Overcoming this challenge is essential to make the method feasible for more complex systems. A way to do that would be to optimize the way weights are set, going beyond heuristics and trial-and-error \cite{weights_1, weights_2, weights_3}. Furthermore, for potentials with degenerate states, the PINN has no way to prefer any particular set of states, and is equally likely to output any linear combination of the degenerate states. This is generally not an issue if we are just looking for solutions of Schr\"odinger's equation for the given potential, since the PINNs will output valid solutions. However, if a particular set of eigenfunctions is preferred, we need to add further physical constraint to the network to ensure the correct output. It is also important to note that for these simple systems the PINN's performances are inferior to those of numerical solvers, both from the point of view of accuracy and speed, if we take into account the training time. For instance, numerical inversion of the Hamiltonian for the particle in a ring leads to errors for the eigenvalues in the order of $10^{-11}$ and can be performed in a few seconds. However, while there are still many challenges with PINNs, especially related to their training, this tool is potentially able to represent any quantum mechanical system given either the governing potential or a sufficient amount of data. Possibly even going beyond what current numerical solvers are able to do. Furthermore, a PINN trained on a more complex system could be utilized to quickly output the wave function $\psi(\vec{x})$ at any point $\vec{x}$, possibly faster then classical solvers.}

\section{Conclusions}
The application and development of deep learning for science and technology has huge potential, and PINNs are among the most promising tools to progress science by using machine learning. Nowadays, one of the most stimulating fields of 
research aims at the development of novel technologies able to actively leverage quantum resources. 
For such purposes, the characterization of the energy spectrum (and the corresponding eigenfunctions) of the system
is essential and it passes through the solution of the Schr\"odinger's equation.  
This tutorial provided the bases to build a PINN to solve the Schr\"odinger's equation in an unsupervised way.
We have discussed the infinite potential well and the particle in a ring, exemplifying systems that allow the reader to fully understand the essential steps to build such type of 
neural networks. 
The research in the use of PINNs in quantum technologies is still relatively novel, requiring more research to find the optimal way to employ such technique. Future directions include finding a better way to optimize the weights, improve on the computation of integrals, building models able to better handle degenerate states, and most importantly applying the method to increasingly more complex systems, \cor{such as atomic nuclei, to assess the role played by coherence in quantum technologies\cite{quantum_coherence}, and to solve stochastic Schr\"odinger's equation modelling decohering processes \cite{Quantum-to-Classical-Coexistence}.}

\vspace{6pt} 




\authorcontributions{Conceptualization, A.M., L.B.; methodology, L.B., A.M., E.P.; software L.B.; writing, L.B., A.M., E.P.}

\funding{The Authors acknowledge the project CQES of the Italian Space Agency (ASI) for having partially supported this research (grant N. 2023-46-HH.0). The authors also acknowledge support from the Qxtreme project funded via the Partenariato Esteso FAIR. }

\dataavailability{The code producing the results in the paper is available from the authors upon reasonable request.}


\conflictsofinterest{The authors declare no conflicts of interest.}

\reftitle{References}


\bibliography{refs}

\PublishersNote{}
\end{document}